\documentclass[graybox]{svmult}

\usepackage{type1cm}        
\usepackage{makeidx}         
\usepackage{graphicx}        
\usepackage{multicol}        
\usepackage[bottom]{footmisc}

\usepackage{newtxtext}       %
\usepackage[varvw]{newtxmath}       
\usepackage[numbers]{natbib}

\makeindex             

\begin{document}

\title*{Ultra-hot Jupiter atmospheres at high spectral resolution}
\titlerunning{Ultra-hot Jupiter atmospheres}
\author{Stefan Pelletier\orcidID{0000-0002-8573-805X}, Daniel Kitzmann\orcidID{0000-0003-4269-3311}, Valentina Vaulato\orcidID{0000-0001-7329-3471}, Ana Rita Costa Silva\orcidID{0000-0003-2245-9579}, Michal Steiner\orcidID{0000-0003-3036-3585} \and David Ehrenreich\orcidID{0000-0001-9704-5405}}
\authorrunning{Pelletier et al.}
\institute{Stefan Pelletier \at Observatoire astronomique de l’Université de Genève, 51 chemin Pegasi 1290 Versoix, Switzerland \email{Stefan.Pelletier@unige.ch}
\and Daniel Kitzmann \at Space Research and Planetary Sciences, Physics Institute, University of Bern, Gesellschaftsstrasse 6, 3012 Bern, Switzerland  \at Center for Space and Habitability, University of Bern, Gesellschaftsstrasse 6, 3012 Bern, Switzerland \email{daniel.kitzmann@unibe.ch}
\and Valentina Vaulato \at Observatoire astronomique de l’Université de Genève, 51 chemin Pegasi 1290 Versoix, Switzerland \email{Valentina.Vaulato@unige.ch}
\and Ana Rita Costa Silva \at Instituto de Astrofísica e Ciências do Espaço, Universidade do Porto, CAUP, Rua das Estrelas, 4150-762 Porto, Portugal  \at Departamento de Física e Astronomia, Faculdade de Ciências, Universidade do Porto, Rua do Campo Alegre, 4169-007 Porto, Portugal \at Observatoire astronomique de l’Université de Genève, 51 chemin Pegasi 1290 Versoix, Switzerland \email{ana.rita@astro.up.pt}
\and Michal Steiner \at Observatoire astronomique de l’Université de Genève, 51 chemin Pegasi 1290 Versoix, Switzerland \email{Michal.Steiner@unige.ch}
\and David Ehrenreich \at Observatoire astronomique de l’Université de Genève, 51 chemin Pegasi 1290 Versoix, Switzerland \email{David.Ehrenreich@unige.ch}
}
%
%
\maketitle
\abstract{
Observations of ultra-hot Jupiters offer an unprecedented opportunity to study the physics of some of the most extreme planetary atmospheres known. 
With exceedingly high amounts of irradiation blasting their upper atmospheres, ultra-hot Jupiters have dayside temperatures comparable to some late type stars enabling refractory metals otherwise condensed in colder planets to exist in the gas phase, all the while still maintaining comparatively cool nightsides.  
The ensuing intense temperature contrasts can give rise not only to strong day-to-night winds, but also to vastly different chemical and cloud properties on opposing hemispheres.  
With its ability to resolve spectral features that are unique to individual chemical species, high resolution spectroscopy can unambiguously disentangle atmospheric signals of exoplanetary origin, which follow a well-defined Keplerian motion, from stationary or pseudo-stationary telluric and stellar lines.
Combined, the high temperature of ultra-hot Jupiters providing access to refractory metals with narrow spectral features and the ability of high-resolution spectroscopy to resolve said narrow lines provides access to a wealth of information about these atmospheres that would otherwise be unavailable at lower resolving powers or for other types of planets.
In this chapter we explore some of the key physical and chemical transitions that differentiate ultra-hot Jupiters from their colder counterparts and highlight the unique opportunities arising from probing their atmospheres using high resolution spectroscopy.
}


\section{Key transitions from hot to ultra-hot Jupiters}
\label{sec:UHJs}

The discovery of 51 Pegasi b~\cite{mayor_jupiter-mass_1995} unveiled the existence of so-called `hot Jupiters', gas giant exoplanets similar to Jupiter in size, but orbiting much closer to their host stars than even Mercury is to the Sun. With orbital periods of a few to several days and being subject to intense stellar irradiation heating their daysides to high temperatures, hot Jupiters are extreme in comparison to the overall exoplanet population and ideally suited for atmospheric characterization. The need for a new even more extreme class of exoplanets, `ultra-hot Jupiters', became apparent following the discovery of KELT-9b~\citep{gaudi_giant_2017}, an exoplanet with an equilibrium temperature above 4000\,K, hotter than some late type stars. Also being gas giants, ultra-hot Jupiters are often defined in the literature either as having an equilibrium temperature (T$_{\mathrm{eq}}$) above $\sim$2000--2200\,K or dayside temperatures above $\sim$2200--2500\,K~\citep{parmentier_thermal_2018, tan_atmospheric_2019, baxter_transition_2020}. But, other than this seemingly arbitrary cutoff in temperature, it is not always clear what physically separates hot from ultra-hot Jupiters. Superficially, these both share common characteristics: short orbital periods, synchronous rotations due to tidal locking, early type host stars, high Keplerian velocities, and elevated dayside temperatures. Indeed, ultra-hot Jupiters are sometimes thought of simply as `hotter hot Jupiters'.  However, the hot to ultra-hot transition coincides with a number of important chemical and dynamical regime changes that make these two types of exoplanets inherently different.

\subsection{Atmospheric puffiness and the dissociation of molecules}
\label{subsec:diss}
\begin{figure}[ht]
\centering
\includegraphics[width=\linewidth]{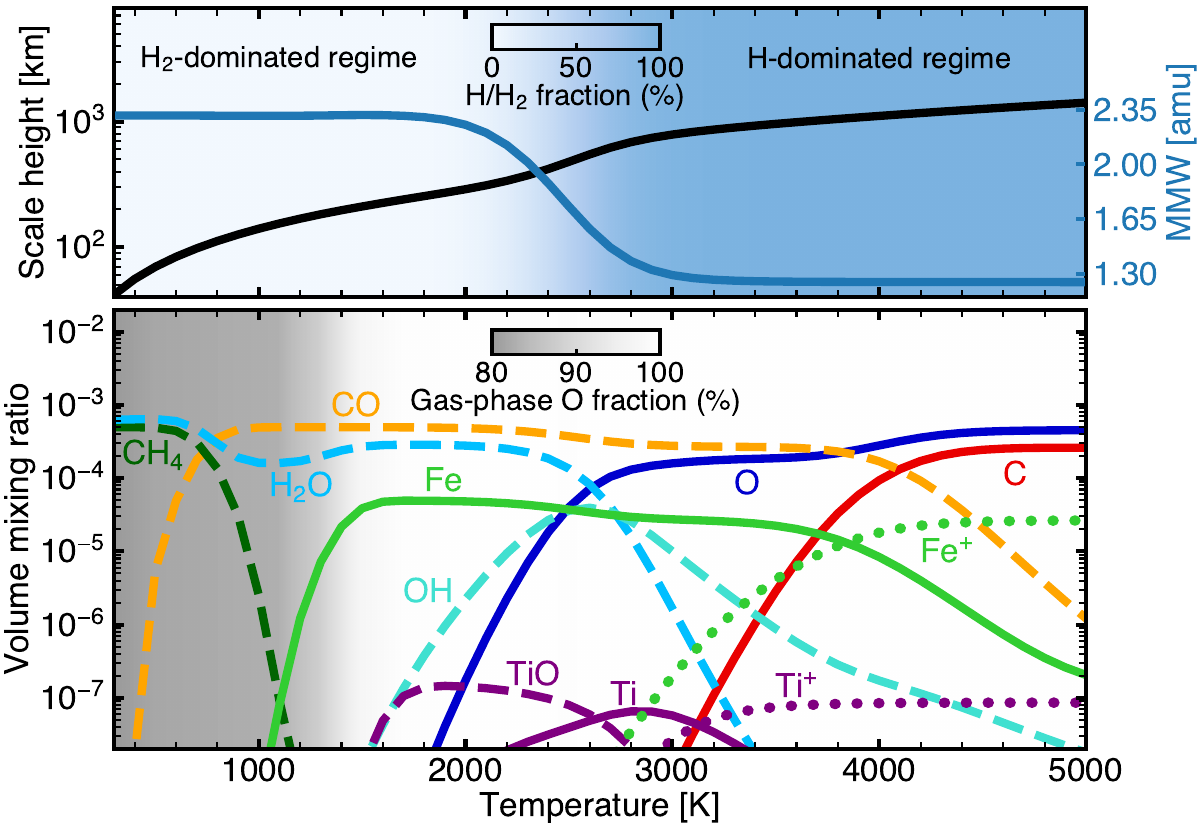}
\caption{Physical and chemical transitions from cold to ultra-hot Jupiters. \textbf{Top:} The atmospheric scale height (solid black line) of a Jupiter-mass planet as a function of temperature at a pressure of 1\,millibar. The secondary axis shows the atmospheric mean molecular weight (MMW, solid blue line) and the blue background shading similarly depicts the fractional dissociation of molecular hydrogen. The atmospheric puffiness steadily increases with temperature and is further increased by the dissociation of molecular hydrogen into atomic hydrogen altering the mean molecular weight.
\textbf{Bottom:} Volume mixing ratios as a function of temperature for important species in gas giant atmospheres predicted by equilibrium chemistry for a solar composition at 1\,millibar using \texttt{FastChemCond}~\citep{stock_fastchem_2018, stock_fastchem_2022, kitzmann_fastchem_2024}. Dashed lines depict molecules, solid lines show neutral elements, and dotted lines represent ions. The grey background shading represents the percentage of the O budget that is missing from the gas phase in colder planets due to the condensation of oxygen-bearing compounds (e.g., MgSiO$_3$), assuming a solar volatile-to-refractory ratio. At low temperatures ($\lesssim$500\,K), most C and O atoms are bound in H$_2$O and CH$_4$ molecules. Beyond $\sim$600\,K, CO begins to replace CH$_4$ as the main C carrier. For typical hot Jupiter atmospheres ($\sim$1000--2000\,K), H$_2$O and CO hold the majority of the gas phase C and O. As the temperature increases above $\sim$2000\,K, H$_2$O begins to thermally dissociate into OH and atomic O. Past $\sim$2500\,K, OH also breaks apart into its atomic components. Although particularly stable owing to its triple bond, even CO molecules dissociate beyond $\sim$3500\,K eventually leaving the full C and O inventory to be in atomic form. With temperatures exceeding $\sim$2000\,K, ultra-hot Jupiters are in a regime where molecular dissociation and ionization are important, where refractory metals like Fe and Ti can exist in the gas phase, and where the full oxygen budget should be accessible to remote sensing.
}
\label{fig:chem}
\end{figure}

As the temperature increases, the thermal dissociation of molecular hydrogen causes the atmosphere to change from being H$_2$ dominated to being atomic H dominated on the dayside of hot giant exoplanets~\citep{tan_atmospheric_2019, bell_increased_2018}. Because gaseous giant planets are primarily made of hydrogen, the change in the atmospheric mean molecular weight ($\mu$) due to a conversion of the main gas component from molecular H$_2$ ($\mu$=2) to atomic H ($\mu$=1) is significant. This compositional transition has the drastic effect of increasing the scale height of the atmosphere (and hence its puffiness) by nearly a factor of 2 (Figure~\ref{fig:chem}, top panel). The hot to ultra-hot Jupiter transition roughly corresponds to when dayside atmospheres go from being H$_2$- to being H-dominated at photospheric pressures. Even for most ultra-hot Jupiters, however, hydrogen will likely still preferentially remain in molecular form on the colder nightside, which can lead to significant day-to-night differences in the atmospheric scale height~\citep{tan_atmospheric_2019, wardenier_decomposing_2021}. The recombination of atomic hydrogen into its molecular form as hot gas is transported around the planet can also decrease the temperature contrast between the day and night sides via the release of latent heat~\citep{bell_increased_2018}.

Similar to H$_2$, another chemical transition that becomes important in the ultra-hot regime is the thermal dissociation of H$_2$O and corresponding production of OH and O~\citep{parmentier_thermal_2018}. As one of the primary sources of opacity in hot gas giant atmospheres in the near-infrared~\citep{gandhi_molecular_2020}, the removal of H$_2$O can significantly change observed atmospheric spectra. The breaking up of water molecules is also particularly important to take into account for abundance inferences, especially when compared relative to other chemical species such as CO, which is more stable and thus less prone to thermal dissociation (Figure~\ref{fig:chem}, bottom panel). Neglecting this effect in ultra-hot Jupiter atmospheres would thus lead to the H$_2$O abundance (and hence the atmospheric O/H) to be underestimated due to unaccounted oxygen held in OH and atomic form. In contrast to the gas phase carbon budget, which can generally be well estimated from the CO abundance alone in the temperature range of most ultra-hot Jupiters~\citep{pelletier_atomic_2025}, obtaining a direct measurement of the complete gas phase oxygen budget will typically require simultaneous constraints on CO, H$_2$O, OH, and O (Figure~\ref{fig:chem}, bottom panel). Alternatively, if only a subset of these can be detected, assumptions of equilibrium chemistry~\citep{ramkumar_new_2025, pelletier_crires_2025, pelletier_enriched_2025} or a parametric abundance profile for H$_2$O~\citep{parmentier_thermal_2018, pelletier_crires_2025, coulombe_broadband_2023}, as well as a given fraction of oxygen that is in atomic form due to dissociation~\citep{brogi_roasting_2023}, are likely necessary to accurately estimate the total O abundance in the gas phase. As a result, accurate O/H inferences for ultra-hot Jupiters can be more challenging relative to hot Jupiters, which are only expected to have CO and H$_2$O as their main O-bearing molecules in a solar composition atmosphere~\citep{line_solar_2021}. Contrastingly, at even higher temperatures ($\gtrsim$4500\,K), the chemistry arguably simplifies, with the full chemical inventory becoming atomic (Figure~\ref{fig:chem}, bottom panel).

\subsection{Release of refractory species to the gas phase}
\label{subsec:ref}
At a given temperature and pressure, the equilibrium state of any chemical species will depend on its condensation temperature~\citep{lodders_solar_2003}. In colder environments, only volatile species that have low condensation temperatures will exist in the gas phase while refractory compounds with higher condensation temperatures will condense into liquid or solid form. One example of this is the upper atmosphere of Jupiter ($T_\mathrm{eq} \sim$ 120\,K), where the only metal-bearing molecules that are in the gaseous state are highly volatile (e.g., CH$_4$, NH$_3$, and the noble gases)~\citep{niemann_galileo_1996}. Also on Earth, our atmosphere can only naturally sustain relatively volatile species (e.g., N$_2$, O$_2$, Ar, and CO$_2$) while rock forming refractory elements (e.g., Mg, Si, and Fe) are in solid state. Meanwhile, H$_2$O molecules can exist in gaseous, liquid, and solid states depending on whether the local atmospheric conditions reach the condensation threshold. Similarly to this, refractory elements that would otherwise be condensed can vaporize and contribute to the overall gas phase budget in higher temperature environments. Extreme examples of this include the photospheres of most stars, where even highly refractory elements (e.g., Ti, Al, and Sc)~\citep{lodders_solar_2003} are vaporized and available to be remotely probed. Going from cold to ultra-hot, exoplanetary atmospheres should therefore show progressive compositional transitions wherein more and more elements of higher refractoriness will contribute to the gas phase with increasing temperature~\citep{fortney_unified_2008}.

Pure equilibrium chemistry will predict refractory species like Fe and TiO to favour the gaseous state around approximately 1400\,K and 1800\,K, respectively, at millibar pressures (Figure~\ref{fig:chem}, bottom panel). For tidally locked gaseous exoplanets, however, the more relevant temperature that will control the release of the species may be that of the nightside rather than the local temperature, if a nightside cold-trap is at play~\citep{parmentier_3d_2013, parmentier_transitions_2016}. As gas is circulated from the hot dayside to the colder nightside, condensates can form and rainout to deeper atmospheric layers which, depending on the advection and mixing timescales, should deplete the atmosphere of certain species at photospheric pressures throughout the planet~\citep{parmentier_3d_2013}. Vertical cold-trapping can also be a mechanism that acts to deplete the upper atmosphere of specific species~\citep{spiegel_can_2009}, similar to how water is cold-trapped out of the stratosphere in Earth's atmosphere. In such scenarios, the observed composition of an exoplanet's atmosphere may greatly differ from what would be expected from chemical equilibrium given the measured temperature on the dayside or terminator due to the rainout of certain species~\citep{kitzmann_fastchem_2024}.

Beyond supplying the refractory metals themselves, the vaporization of condensed species also releases the fraction of oxygen atoms removed from the gas phase by the formation of refractory compounds (e.g., MgSiO$_3$, Mg$_2$SiO$_4$, and Fe$_2$O$_3$).  Indeed, hot Jupiters and colder planets that do not have their full refractory inventory in gas form in their atmosphere will have a portion of their oxygen budget missing, inaccessible to remote sensing. While this accounts for roughly 20\% of the O-budget for a solar composition gas (see background shading of Figure~\ref{fig:chem}, bottom panel), the contribution can be higher. For instance, if the planet has a sub-solar volatile-to-refractory ratio, more refractories would be available to sequester O atoms from the gas phase~\citep{turrini_tracing_2021, lothringer_new_2021, pacetti_chemical_2022, chachan_breaking_2023, fonte_oxygen_2023}. The uncertain fraction of oxygen locked away in condensates serves as a fundamental caveat to how well any measured atmospheric O/H ratio truly reflects the composition of the bulk envelope~\citep{pelletier_where_2021}. Ultra-hot planets with temperatures elevated enough to prevent any condensates from forming anywhere throughout their atmospheres can therefore provide access to the bulk envelope composition directly from the gas phase.

\begin{figure}[hb]
\centering
\includegraphics[width=\linewidth]{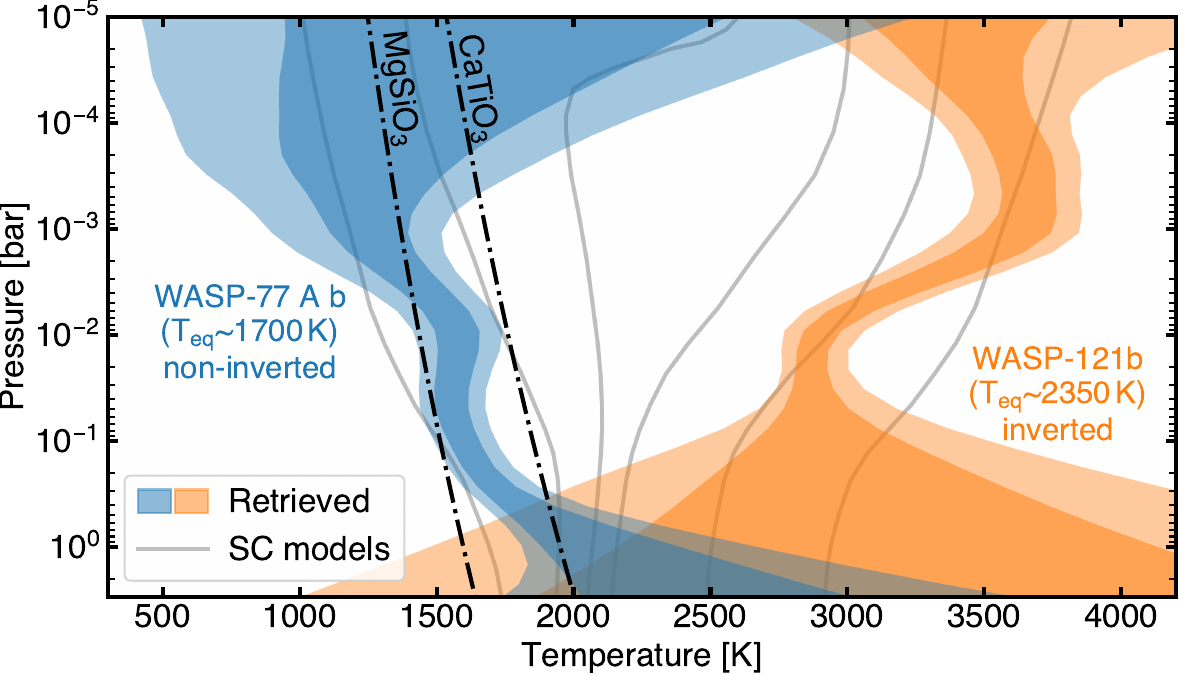}
\caption{Vertical temperature structure transition from hot to ultra-hot Jupiters.  Shown in colour are the retrieved dayside temperature-pressure (TP) profiles (one and two sigma contours) for IGRINS observations~\citep{line_solar_2021, smith_combined_2024, smith_roasting_2024, bazinet_subsolar_2024} of the hot Jupiter WASP-77 A b (blue) and 
of the ultra-hot Jupiter WASP-121b (orange) compared to a grid of one dimensional self-consistent (SC) radiative-convective-thermochemical equilibrium atmosphere models generated using the Sc-CHIMERA code~\citep{mansfield_unique_2021}
(solid grey lines). WASP-77 A b lies well within the non-inverted hot Jupiter regime while WASP-121b shows a strong thermal inversion characteristic of ultra-hot Jupiters.  The onset of stratospheres in hot giant planets is thought to be the result of refractory species that have strong optical opacity (e.g., TiO, VO), condensed at lower temperature, being released into the gas phase. 
The dot dashed black lines show the 50\% condensation curves for MgSiO$_3$ and CaTiO$_3$ assuming a solar metallicity, to the left of which these species energetically prefer being in a condensed state.
Although the retrieved dayside TP profile for WASP-121b is significantly hotter, it may still be it crosses the condensation curve of CaTiO$_3$ on the nightside, causing some fraction of the titanium budget to be removed from the gas phase and cold-trapped to deeper atmospheric layers~\citep{pelletier_enriched_2025, parmentier_3d_2013}.
}
\label{fig:TP_transition} 
\end{figure}

\subsection{Optical absorbers and the driving of thermal inversions}
\label{subsec:inv}
Another consequence of the release of refractory atoms and their corresponding oxides and hydrides to the gas phase is the addition of numerous new opacity sources to the atmosphere.  Notably, many of these have strong spectral lines (e.g., Fe, Ca, and Cr) and bands (e.g., TiO and VO) concentrated at optical wavelengths where most of the stellar spectral energy distribution is emitted, in contrast to typical molecules (e.g., H$_2$O, CO, and CO$_2$) that preferentially show features in the infrared. The introduction of optical absorbers to a planetary atmosphere can therefore lower the pressure at which the stellar energy is deposited, adding heat to the upper atmosphere. This is analogous to the ozone layer on Earth which efficiently absorbs UV sunlight causing the temperature to go from decreasing with altitude in the troposphere to increasing with altitude in the stratosphere. For highly irradiated exoplanets, the presence of strong optical absorbers can drastically change the thermal structure by driving an atmospheric inversion~\citep{fortney_unified_2008, hubeny_possible_2003}. 
This has indeed been observed for exoplanets using high resolution spectroscopy, with ultra-hot Jupiters ubiquitously showing thermally inverted thermal structures~\citep{petz_pepsi_2025} and typical hot Jupiters contrastingly having non-inverted temperature-pressure profiles (Figure~\ref{fig:TP_transition}). In many cases, however, it remains unclear which optical absorber (e.g., TiO, VO, H$^{-}$, or a combination of other metals/molecules) is responsible for the observed temperature inversions~\citep{spiegel_can_2009, lothringer_extremely_2018, gandhi_h-_2020, piette_temperature_2020}. 
A change from a negative to a positive lapse rate will fundamentally change the thermal spectrum of an exoplanet to showing emission rather than absorption lines. 
Recent observations suggest that WASP-122b (T$_{\mathrm{eq}} \sim$ 1950\,K) may be near the non-inverted to inverted transition~\citep{panwar_roasting_2025}.

\subsection{Ionization and the supply of free electrons}
\label{subsec:ionization}
In high temperature atmospheres, the ionization of atoms is expected to become important, especially in low pressure regions. For ultra-hot Jupiters, this is due to both thermal processes, amplified by the presence of a stratosphere, and photoionization~\citep{kitzmann_peculiar_2018, fossati_data-driven_2020}. The ionization of atoms also has the effect of adding free electrons to the atmosphere which, in addition to changing the mean molecular weight, has numerous implications that are characteristic of ultra-hot Jupiters.  For one, the partial ionization of the atmosphere can result in a coupling of the gas to any existing planetary magnetic field, adding a drag force that can hinder the transport of heat to the nightside~\citep{perna_magnetic_2010, batygin_inflating_2010, rogers_magnetic_2014, arcangeli_climate_2019}. The addition of free electrons to the atmosphere, in tandem with the presence of atomic hydrogen, will also result in H$^{-}$ bound-free and free-free interactions, which are important continuum absorbers at high temperatures~\citep{bell_free-free_1987, john_continuous_1988}.  While clouds, hazes, and collision induced absorption can set the opacity floor in spectra of colder giant planets~\citep{sing_continuum_2016}, in ultra-hot Jupiter atmospheres this is rather expected to be set by H$^{-}$~\citep{coulombe_broadband_2023, lothringer_extremely_2018}. More specifically, H$^{-}$ bound-free absorption can dominate the spectrum below $\sim$1.6\,$\mu$m, masking spectral features from species such as H$_2$O, TiO, and VO while H$^{-}$ free-free absorption provides some added continuum opacity mainly at longer wavelengths~\citep{arcangeli_h-_2018}. 

The significant ionization of metals at low atmospheric pressures will result in highly non-uniform abundance profiles for atomic species and their ions, which (similar to molecular dissociation) can complicate any modelling and physical property inference attempts.  In particular, atmospheric retrievals that assume constant with altitude abundance profiles are likely no longer a valid approximation for ultra-hot Jupiters as they would be for hot Jupiters and can lead to biases in inferred abundances~\citep{pelletier_enriched_2025}.  This is especially true for elements with weakly bound outer electron shells that ionize readily, such as alkali metals.
However, the combined presence of neutral atoms and ions together 
can provide a useful chemical thermometer, wherein their relative proportions at any given pressure should be related to the temperature via the Saha equation~\citep{saha_liii_1920}.




\begin{svgraybox}
The transition from hot to ultra-hot Jupiters corresponds to numerous important physical and chemical changes, with dayside atmospheres becoming H dominated, with H$_2$O dissociation forming OH, with condensed refractories being vaporized, with thermal structures becoming inverted, and with atoms significantly ionizing and driving new continuum opacity sources. All in all, ultra-hot Jupiters should not be thought of as simply hotter hot Jupiters, but as a class of their own with different physical processes governing their atmospheres.
\end{svgraybox}

\section{Opportunities arising from high-resolution observations of ultra-hot Jupiter atmospheres}
\label{sec:UHJ_at_HR}

\begin{figure}[ht]
\centering
\includegraphics[width=\linewidth]{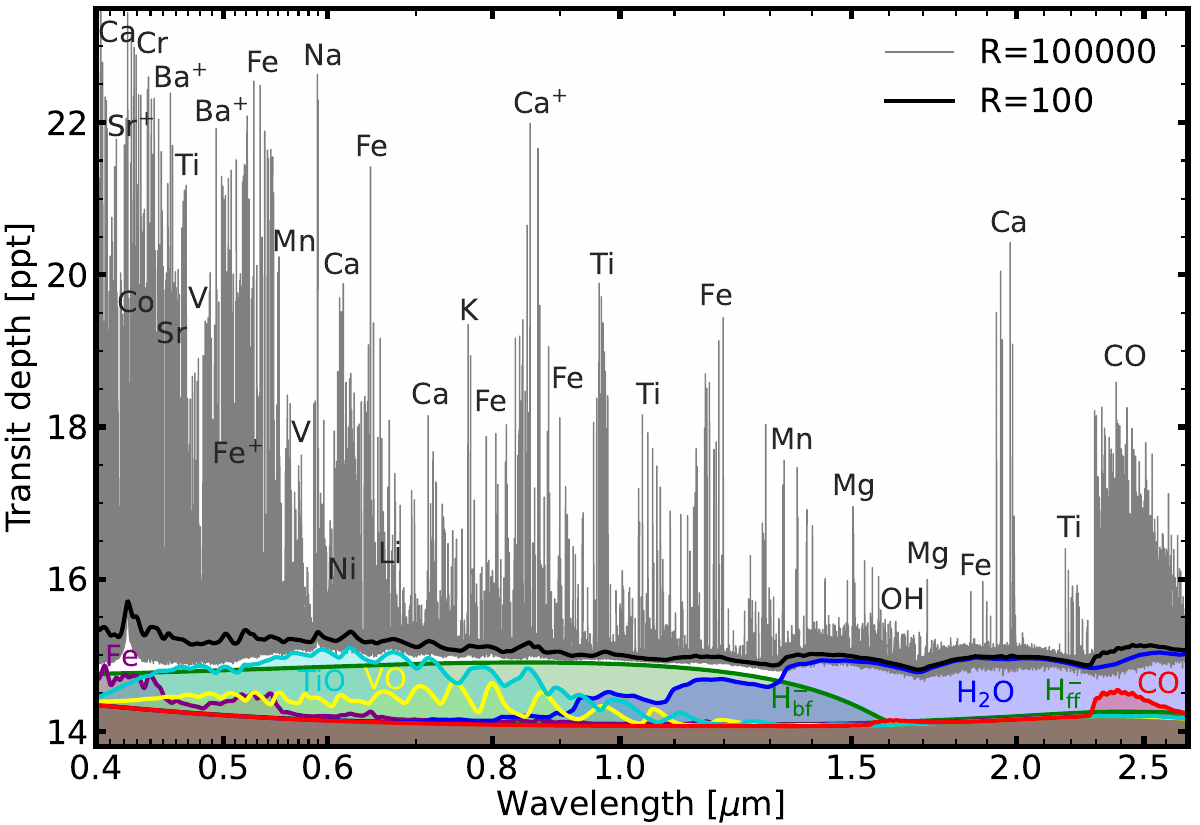}
\caption{
Contribution of chemical species at low and high spectral resolution in a typical ultra-hot Jupiter transmission spectrum. The spectrum containing all opacity sources at a spectral resolution $R$ = 100,000 is shown in grey, while the thick solid lines show important broadband absorbers, binned at $R$ = 100. The atmospheric template assumes the parameters~\citep{bourrier_optical_2020} and composition~\citep{pelletier_crires_2025} of a WASP-121b-like planet in chemical equilibrium at 3000\,K.  Included in the model are important optical and near-infrared opacity sources from species either expected~\citep{kitzmann_fastchem_2024, kitzmann_mantis_2023} or previously detected~\citep{smith_roasting_2024, prinoth_titanium_2025} in its atmosphere. While molecules such as TiO, VO, H$_2$O, and CO have rotational-vibrational transitions that form broad bands detectable at lower spectral resolution, atoms only have line transitions that, for the most part, cannot be probed without higher resolving powers.
}
\label{fig:transit_spec}
\end{figure}

The previous section discusses some of the transitions that are expected to occur from hot to ultra-hot gas giant exoplanets. However, most of these, albeit well based on our understanding of physics and chemistry, have until recently remained only theoretical predictions. Although a planet's spectrum is shaped by the physics and chemistry governing its atmosphere, part of the difficulty arises because much of that information is lost at low spectral resolving powers.  On the other hand, high resolution observations can access more chemical species and lower atmospheric pressures by probing narrow spectral features (Figure~\ref{fig:transit_spec}) and therefore provide a unique opportunity to probe extreme atmospheres. We refer the reader to Section 4.2 of Chapter 15 (\textit{Observational and computational characterization of exoplanet atmospheres}) or to refs.~\cite{birkby_spectroscopic_2018, snellen_exoplanet_2025} for an overview of the high resolution spectroscopy method in the context of exoplanet atmospheres. We now discuss some of the key advancements made in recent years from observations of ultra-hot Jupiter atmospheres at high spectral resolution, many of which were pioneered by studies involving members of the National Centre of Competence in Research PlanetS network.

\subsection{Access to refractory elements}

Arguably one of the biggest recent advancements in the study of ultra-hot Jupiter atmospheres is the prediction~\citep{kitzmann_peculiar_2018} and observed presence~\citep{hoeijmakers_atomic_2018, hoeijmakers_spectral_2019} of neutral and ionized alkaline earth and transition metals. Previously, high resolution observations of exoplanets had mostly focused on molecules in the infrared or alkali metals in the optical, species that had been also observable with space-based observatories such as HST and Spitzer. In recent years, high resolution observations of ultra-hot Jupiters have led to a wealth of new elements being detected, greatly increasing the amount of chemical species now detected in exoplanetary atmospheres~\citep{prinoth_titanium_2025, hoeijmakers_spectral_2019, pelletier_vanadium_2023, borsa_high-resolution_2022, merritt_inventory_2021, gibson_relative_2022, prinoth_titanium_2022, silva_detection_2022, kesseli_atomic_2022, prinoth_time-resolved_2023, simonnin_time-resolved_2025, hoeijmakers_mantis_2024, nugroho_high-resolution_2017, cont_detection_2021, cont_silicon_2022, cont_atmospheric_2022}. With many of these species having different line strengths, condensation temperatures, and ionization thresholds, this has widely expanded the range of physics accessible to be probed in ultra-hot Jupiter atmospheres. Even now in the JWST era, the remote sensing of refractory metals and ions is an area where high resolution spectroscopy remains unmatched.

\subsubsection{An indirect probe into cold traps and nightside cloud compositions}
\begin{figure}[!htbp]
\centering
\includegraphics[width=0.975\linewidth]{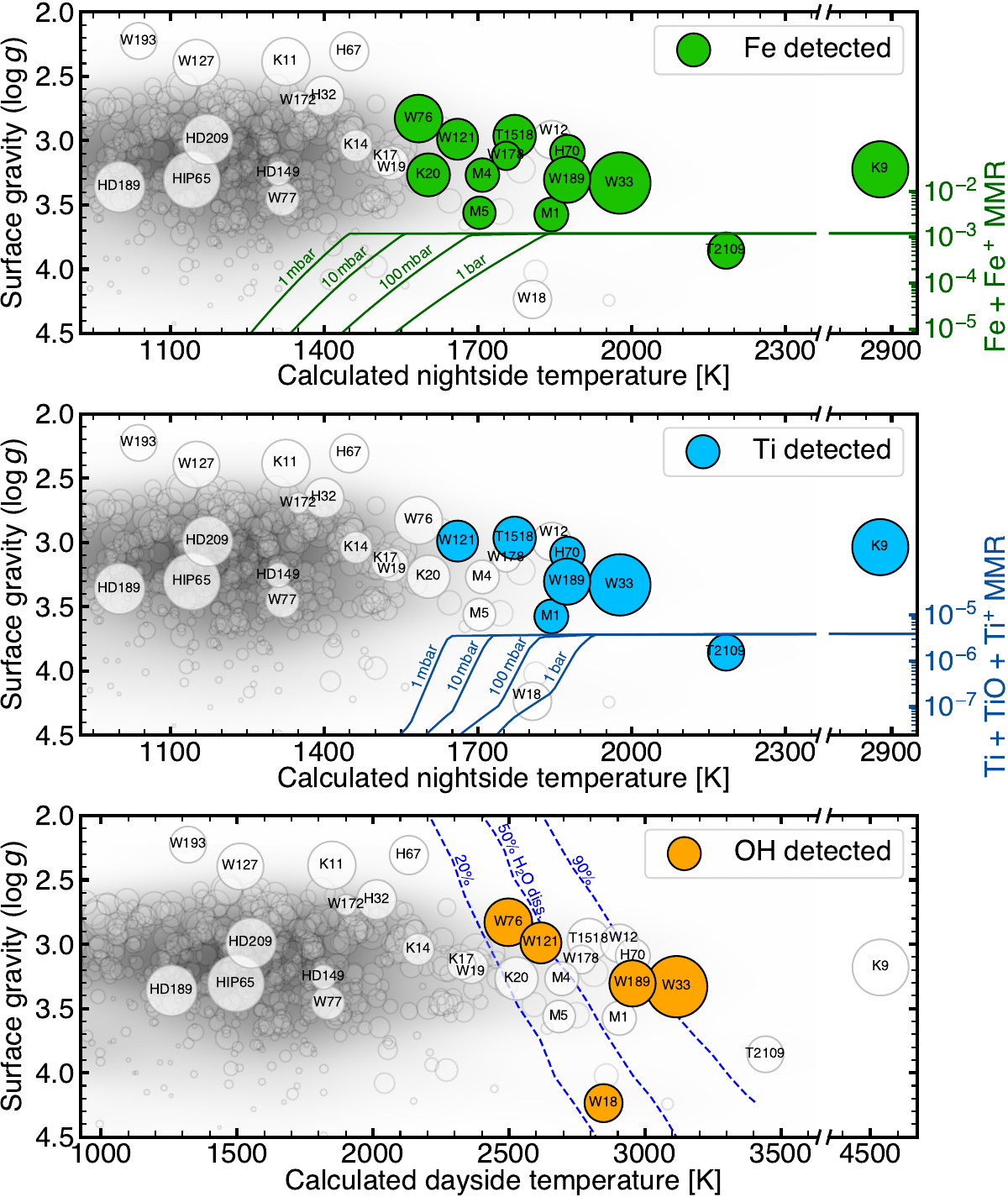}
\caption{
Fe, Ti, and OH high-resolution significant detections in the literature compared to the overall exoplanet population (grey). Day and nightside temperatures are estimated assuming the heat redistribution relation of ref.~\cite{parmentier_cloudy_2021} for the case of nightside clouds. Circle sizes correspond to observational favourability. For the planet names, `W'= WASP, `K' = KELT, `H' = HAT-P, `T' = TOI, and `M' = MASCARA. 
\textbf{Top:} Reported detections of Fe (green) as a function of nightside temperature. The secondary axis shows the total mass mixing ratio (MMR) of Fe and Fe$^+$ predicted by \texttt{FastChemCond}~\citep{kitzmann_fastchem_2024} for a solar-composition gas at different pressure levels. While WASP-76b is currently the lowest temperature ultra-hot Jupiter with Fe unambiguously detected, it shows no sign of depletion~\citep{pelletier_vanadium_2023, gandhi_retrieval_2023}, suggesting that the onset of Fe in planetary atmospheres may occur at lower temperatures.
\textbf{Middle:} Same as middle, but for Ti (blue). Although Ti has been detected on WASP-121b it was found to be underabundant~\citep{pelletier_enriched_2025, prinoth_titanium_2025}, suggesting that this planet may border the gaseous onset of titanium species in ultra-hot Jupiter atmospheres.
\textbf{Bottom:} Reported detections of OH (orange), now as a function of the dayside temperature. The dashed blue lines mark contours of where H$_2$O thermal dissociation reaches 20\%, 50\%, or 90\% at the 1.4\,$\mu$m photosphere~\citep{parmentier_thermal_2018}. All OH detections fall in the region of the parameter space where H$_2$O should be at least 20\% dissociated.
}
\label{fig:OH_Fe_Ti}
\end{figure}

With its high natural abundance in the universe and relatively high density of strong spectral features in the optical, Fe is one of the most readily observable chemical species in ultra-hot Jupiter atmospheres. Indeed, neutral atomic Fe can be detected using high resolution spectroscopy even from a few hours of observing time on 2-m class ground-based telescopes~\citep{bello-arufe_exoplanet_2022, lowson_multiepoch_2023, borsato_small_2024}. So far, Fe has been detected in all giant planets with temperatures equal to or above that of WASP-76b (T$_{\mathrm{eq}} \sim$ 2200\,K), for which high resolution, high signal-to-noise ratio optical observations have been published (Figure~\ref{fig:OH_Fe_Ti}, top panel). Contrastingly, no significant detections of Fe have yet been made in any planet below this temperature threshold. However, we caution that this may be due to a coincidental lower observational favourability of known targets that are only slightly colder than WASP-76b. So far, non-detections of Fe from high sensitivity data sets have only been reported for WASP-19b (T$_{\mathrm{eq}} \sim$ 2100\,K)~\citep{sedaghati_spectral_2021}, KELT-17b (T$_{\mathrm{eq}} \sim$ 2100\,K)~\citep{stangret_high-resolution_2022}, and HAT-P-67b (T$_{\mathrm{eq}} \sim$ 1900\,K)~\citep{bello-arufe_transmission_2023}.  Meanwhile, no ultra-hot Jupiters for which Fe abundance measurements exist show any definitive signs of depletion on their daysides or terminators~\citep{pelletier_crires_2025, smith_roasting_2024, pelletier_vanadium_2023, gibson_relative_2022, gandhi_retrieval_2023, maguire_high-resolution_2023}, even if some rainout may be occurring on the nightside~\citep{ehrenreich_nightside_2020}.

If Fe is fully in the gas phase on WASP-76b and hotter planets, but at least partially depleted on WASP-19b allowing it to remain undetected, then this would place the onset of gaseous Fe in giant planetary atmospheres approximately in the T$_{\mathrm{eq}}$ = 2100--2200\,K range.  While one might expect Fe to be present in the gas phase well below this temperature threshold given its 50\% condensation temperature being around 1350\,K at 10$^{-4}$ bar~\citep{lodders_solar_2003}, the more relevant temperature to consider may be the colder nightside where condensation could still occur.  If a cold-trap mechanism is at play, gaseous Fe from the dayside that is circulated to the nightside may rainout without ever being able to circulate back to the upper atmosphere (see Section~\ref{subsec:ref}). Instead considering the calculated nightside temperatures~\citep{parmentier_cloudy_2021}, the threshold for Fe being released from being cold trapped occurs at nightside temperatures around 1550\,K (Figure~\ref{fig:OH_Fe_Ti}, top panel).  Comparing this to the expected equilibrium condensation of Fe~\citep{kitzmann_fastchem_2024}, then this could imply that the composition of the terminator is set by the nightside composition at around the 10\,mbar level. However, we note that the process of condensing, mixing, and revaporization of Fe and other species which will ultimately set their abundances at the dayside or terminator regions of the atmosphere will depend on condensation microphysics and advection timescales and thus may not be well approximated by a pure equilibrium chemistry calculation~\citep{parmentier_3d_2013, spiegel_can_2009, powell_formation_2018}.

Ultimately, more studies investigating the presence or absence (and depletion degree) of Fe in planets colder than WASP-76b will be necessary to better trace out and understand the onset of gaseous Fe in giant planetary atmospheres and whether it matches the onset of thermal inversions. In particular, if the release of species to the gas phase depends on countering gravitational settling and mixing gas from below the photosphere, then this may be strongly dependent on the surface gravity as well.  For example, although Fe was not seen in the atmosphere of the super puffy HAT-P-67b, ionized calcium (which has a higher condensation temperature than Fe~\citep{lodders_solar_2003}) was detected~\citep{bello-arufe_transmission_2023}. Fe was also tentatively observed on the bloated giant hot Jupiter WASP-172b (T$_{\mathrm{eq}} \sim$ 1750\,K)~\citep{seidel_detection_2023-1}, potentially suggesting that low gravity planets can more easily break cold-traps and circulate metals to the upper atmosphere.

The same logic can be applied to titanium, an element that is also relatively easy to detect but which has a higher condensation temperature compared to Fe. If species are indeed released to the gas phase from being condensed, it is therefore likely that there are planets where Fe is in the gas phase, but not Ti. A compilation of all reported significant and unambiguous detections of Ti in the literature\footnote{We specifically do not include any claimed detections of TiO as these can be ambiguous at low resolution and suffer from known line list issues at high resolution~\citep{prinoth_titanium_2022, hoeijmakers_search_2015, mckemmish_exomol_2019}} does appear to indicate that the onset of titanium in giant planetary atmospheres may occur at higher temperatures than for Fe (Figure~\ref{fig:OH_Fe_Ti}, middle panel).  Most interestingly, the coldest planet for which Ti has been unambiguously detected (WASP-121b) shows a signal that is significantly weaker than expected by models and which was only found due to the extreme sensitivity of the ESPRESSO data in 4-UT mode~\citep{pepe_espresso_2021, borsa_atmospheric_2021, seidel_detection_2023, seidel_vertical_2025}. 
This could suggest that while Ti is present in the gas phase, some of it is still missing, which could explain the previous non-detections and abundance upper limits that had implied the presence of a nightside cold-trap~\citep{pelletier_crires_2025, smith_roasting_2024, gibson_relative_2022, hoeijmakers_mantis_2024, gandhi_retrieval_2023, maguire_high-resolution_2023, hoeijmakers_hot_2020, merritt_non-detection_2020}. 
Ti was also found to be severely depleted in the atmosphere of WASP-76b (T$_{\mathrm{eq}} \sim$ 2200\,K)~\citep{pelletier_vanadium_2023, gandhi_retrieval_2023}, which suggest that WASP-121b (T$_{\mathrm{eq}} \sim$ 2350\,K) is at the edge of the titanium shoreline in giant exoplanet atmospheres.  Similar to the case of Fe, this would indicate that the abundance of refractories in ultra-hot Jupiters may be approximately represented by the equilibrium chemistry composition of the nightside at around 10\,mbar (Figure~\ref{fig:OH_Fe_Ti}, middle panel).

We stress that this is still speculative and more observations are needed to explore and confirm the onset of metals in hot and ultra-hot exoplanets. For example, a detection of Ti and tentative detection of Fe were reported on the hot Jupiter HD~149026~b~\citep{ishizuka_neutral_2021}, which is much cooler than all other planets with robust detections of these species. However, follow up observations were unable to recover any similar signals, despite their better data sensitivity~\citep{biassoni_high-resolution_2024}. To gain a more complete understanding of the ongoing vaporization of metals in hot Jupiters, precise relative abundance inferences of species spanning a range of condensation temperatures will be necessary across the population of giant exoplanets amenable for atmospheric characterization.
Although we focused here on Fe and Ti as they are major opacity contributors easily detectable with high resolution spectroscopy, similar investigations can be made for other elements as well. Indeed, probing the presence or absence of species with different condensation temperatures can serve as a indirect probe into the thermal structure and composition of clouds on the nightside.

\subsubsection{CO and a window into bulk envelope compositions}

Carbon monoxide is an important carrier of both oxygen and carbon in typical hot and ultra-hot Jupiter atmospheres. This makes CO a critical species to constrain in order to measure both the C/H and O/H elemental abundance ratios. However, CO has fewer spectral bands compared to H$_2$O and CH$_4$, which are not as strong as those of e.g., CO$_2$, making CO relatively difficult to unambiguously identify at lower spectral resolving powers.  Even with JWST, CO only provides some excess opacity near 2.3 and 4.6 microns which can overlap with different absorbers and often leads to degenerate abundance inferences~\citep{lothringer_refractory_2025}. On the other hand, at high spectral resolution CO is one of the most commonly detected molecules in exoplanetary atmospheres~\citep{line_solar_2021, pelletier_where_2021, brogi_signature_2012, brogi_detection_2013, brogi_carbon_2014, giacobbe_five_2021, carleo_gaps_2022, guilluy_gaps_2022, vansluijs_carbon_2023, choi_early_2025}. The simultaneous overlap of CO lines with those of H$_2$O and some metals can also allow their relative abundances to be much better constrained then their respective absolute abundances~\citep{brogi_roasting_2023, smith_roasting_2024}.

CO is also expected to be the primary carbon-bearing species in gas giant planets for a wide range of temperatures (Figure~\ref{fig:chem}, bottom panel).  Its strong triple bond makes it resistant to thermal dissociation, meaning that it will not be as sensitive to model assumptions on its abundance profile compared to H$_2$O or even metals like Fe~\citep{pelletier_crires_2025, pelletier_enriched_2025}.  CO is also expected to be relatively constant throughout the atmosphere of hot and ultra-hot Jupiters (both day and night sides), making CO an excellent benchmark to compare other molecules to~\citep{savel_diagnosing_2023}. Obtaining abundance ratios relative to CO such as CO/H$_2$O, CO/Fe, and CO/Ti for a wide range of ultra-hot Jupiters spanning different irradiation levels and surface gravities will be especially insightful in understanding chemical transitions related to dissociation and condensation expected to occur in their atmospheres. Likely only on the hottest known exoplanets with fully dissociated atmospheres such as KELT-9b does CO not play a dominant role in the metallic budget, at which point the bulk composition requires direct constraint on atomic C and O~\citep{pelletier_atomic_2025}.

With the vaporization of all condensed compounds in the hottest giant planets, the bulk chemical inventory of a planetary envelope becomes fully accessible via the gas phase.
This is of particular interest from a planet formation and evolution point of view, wherein the bulk envelope composition of the primordially accreted atmosphere may retain information about its accretion history~\citep{mordasini_imprint_2016}. The simultaneous access provided by ultra-hot Jupiters to not only C and O, but also multiple rock forming elements can shed light into the ice-to-rock ratio of the material from which giant planet are formed~\citep{pelletier_crires_2025, lothringer_new_2021, chachan_breaking_2023, smith_roasting_2024, chachan_strong_2025}. The comparison of elemental abundance ratios in the planet to those of the star can also test the hypothesis of whether planets preserve certain stellar abundance ratios such as Mg/Fe, as often assumed in interior modelling of small planets~\citep{cadieux_new_2024}.

\subsection{A direct probe into thermal dissociation and ionization}
While chemical equilibrium predicts that the thermal dissociation of molecules will occur at high temperatures (Figure~\ref{fig:chem}, bottom panel), the observational confirmation of its importance in exoplanet atmospheres has remained elusive. At low resolution spectroscopy, the best previous evidence for molecular dissociation in exoplanet atmospheres was indirect, via the effect of H$^{-}$ on observed spectra~\citep{coulombe_broadband_2023, arcangeli_h-_2018}. In the case of H$_2$O, the uncertainty regarding its degree of dissociation can lead to ambiguous abundance inferences, wherein low values relative to other species can either be due to the atmosphere actually being oxygen (water) poor or due to H$_2$O only appearing underabundant due to it being partially dissociated. Without making underlying assumptions on the chemistry, these scenarios are difficult to distinguish~\citep{pelletier_enriched_2025} and can bias the inferred C/O ratio to higher values~\citep{madhusudhan_carbon-rich_2011}, which can have major implications from a planet formation and evolution standpoint~\citep{oberg_effects_2011}.

High-resolution observations of ultra-hot Jupiters have recently provided direct access to OH~\citep{brogi_roasting_2023, smith_roasting_2024, weiner_mansfield_metallicity_2024, nugroho_first_2021, landman_detection_2021, bazinet_quantifying_2025}, which is a byproduct of H$_2$O dissociation.  While OH is difficult to observe at low spectral resolution as it does not have distinct strong molecular bands, its forest of narrow lines can readily be detected using high resolution spectroscopy.  Probing OH not only offers direct evidence that dissociation is crucial in ultra-hot Jupiter atmospheres, but also--when its abundance is measured alongside that of H$_2$O--can reveal the water dissociation fraction~\citep{bazinet_quantifying_2025, gandhi_revealing_2024}, enabling a better inference of the O/H and C/O ratios~\citep{brogi_roasting_2023, smith_roasting_2024}. The simultaneous access to OH and H$_2$O also provides a chemical thermometer, wherein their relative abundance predicted from chemical equilibrium is strongly dependent on temperature and pressure~\citep{kitzmann_peculiar_2018, wright_spectroscopic_2023}.  While all of the planets for which OH has been unambiguously detected lie in the expected parameter space (Figure~\ref{fig:OH_Fe_Ti}, bottom panel), future works measuring OH abundances will enable us to better test theoretical predictions of the thermal dissociation of molecules from equilibrium chemistry~\citep{parmentier_thermal_2018, kitzmann_peculiar_2018, gandhi_revealing_2024}. For example, the presence of OH on colder planets, as tentatively observed on WASP-127b (T$_{\mathrm{eq}} \sim$ 1400\,K)~\citep{boucher_co_2023}, may also reveal the importance of photodissociation of H$_2$O and the production of OH in the upper atmospheres of colder exoplanets.

High resolution observations of ultra-hot Jupiters have also unlocked access to numerous ions that have been detected in their atmospheres~\citep{hoeijmakers_atomic_2018, hoeijmakers_spectral_2019, merritt_inventory_2021, prinoth_titanium_2022, silva_detection_2022, sing_hubble_2019, hoeijmakers_high-resolution_2020}. Being each short of one electron, these ions are the source for the free electrons whose interactions set the continuum in ultra-hot gaseous atmospheres. A direct measure of the ionization fraction can therefore help break degeneracies regarding the absolute abundances and the opacity floor pressure level~\citep{benneke_atmospheric_2012, vaulato_hydride_2025, vaulato_atmospheric_2025}. As the relative proportion of an element in neutral and ionized state will depend on the temperature, simultaneous abundance constraints on these (e.g., Fe and Fe$^{+}$) additionally serve as chemical thermometers to help better constrain the thermal structure of ultra-hot Jupiter atmospheres. The highly non-uniform abundance profiles of ions also provides a unique window into the upper layers of an atmosphere, unlike the neutral states that can probe a wide range of pressures~\citep{kesseli_up_2024}.

\subsection{Measuring atmospheric inversions from line shapes}
High resolution spectroscopy observations of the thermal emission of exoplanets are also ideally suited for resolving the shapes of spectral lines to determine whether or not an atmosphere has a stratosphere and giving insights into planetary dynamics and rotation~\citep{bazinet_subsolar_2024, lesjak_retrieval_2023, silva_espresso_2024, cont_retrieving_2025, guilluy_gaps_2025}.  
While this can be ambiguous at lower spectral resolving powers~\cite{mansfield_unique_2021, arcangeli_h-_2018, haynes_spectroscopic_2015} due to degeneracies between composition and thermal profile, at high spectral resolution the resolving of spectral features can unambiguously distinguish between emission and absorption line profiles~\citep{hoeijmakers_mantis_2024, schwarz_evidence_2015, pino_neutral_2020, pino_gaps_2022}.  
In particular, with lower pressures probing dynamically distinct regions of the atmosphere~\citep{lee_mantis_2022}, the origin of inversions originating from the upper atmosphere, as models would predict (Figure~\ref{fig:TP_transition}), or from higher pressures due to partial mixing of cold-trapped refractory elements such as TiO, can be better characterized. Not only that, but high-resolution spectroscopy can directly connect thermal inversions with the optical absorbers that drive such mechanism, being sensitive to TiO, VO, and metals~\cite{pelletier_vanadium_2023, prinoth_titanium_2022, prinoth_time-resolved_2023, maguire_high_2024}. Interestingly, while TiO has long been thought to be the main driver of thermal inversions in hot giant exoplanets, its presence cannot always be confirmed, and is often found to be underabundant~\cite{pelletier_enriched_2025, prinoth_titanium_2025, pelletier_vanadium_2023, gibson_relative_2022, gandhi_retrieval_2023, maguire_high-resolution_2023} even when a strong thermal inversion exists~\citep{pelletier_crires_2025, smith_roasting_2024}. More observations of hot and ultra-hot Jupiters spanning a range of equilibrium temperatures will be particularly well suited for further investigating the transition from non-inverted to inverted thermal structures~\citep{petz_pepsi_2025} and how it compares to general circulation model predictions~\citep{roth_hot_2024}.

\subsection{Insights into dynamics and phase-resolved asymmetries}

High resolution observations of ultra-hot Jupiters have also provided the first phase-resolved study of a significant asymmetry between the morning and evening terminators of an exoplanet~\citep{ehrenreich_nightside_2020}.  This is made possible by the ability of high resolution observations to spectroscopically resolve the different velocity shifts of opposing atmospheric limbs of transiting exoplanets. In addition, the puffy dayside atmospheres of ultra-hot Jupiters produce high amplitude signals that can allow their 3D shapes as well as any asymmetries between their different limbs to be probed via both metals and molecules~\citep{pelletier_vanadium_2023, kesseli_atomic_2022, gandhi_retrieval_2023, ehrenreich_nightside_2020, guilluy_gaps_2025, gandhi_spatially_2022, kesseli_confirmation_2021, van_sluijs_assessing_2025}. This has provided a new avenue to study the rainout and dissociation of chemical species from the day to the nightside~\citep{ehrenreich_nightside_2020, wardenier_phase-resolving_2024}, as well as any temperature or cloud asymmetries between opposing limbs~\citep{wardenier_decomposing_2021, savel_diagnosing_2023, wardenier_modelling_2023, wardenier_phase-resolving_2024, wardenier_pretransit_2025, savel_no_2022, beltz_magnetic_2022, beltz_magnetic_2023, beltz_comparative_2024}.

\begin{svgraybox}
So far, observations of ultra-hot Jupiter atmospheres at high spectral resolution have been a gift that keeps on giving in terms of wealth of information and new discoveries, providing an unprecedented opportunity at better understanding the chemical and physical processes that govern their atmospheres and correspondingly improving our models.
Although ultra-hot Jupiters are an extreme case in terms of signal amplitude, as our methods and observational facilities improve, they can provide a taste of what we will hopefully one day be able to do on smaller and colder planets in the era of the ELTs and beyond.

\end{svgraybox}






\begin{acknowledgement}
S.P.\ thanks Michael Line for helpful discussions on the topic. 
This project has been carried out within the framework of the National Centre of Competence in Research PlanetS supported by the Swiss National Science Foundation under grant 51NF40\_205606. The authors acknowledge the financial support of the SNSF.
\end{acknowledgement}
\ethics{Competing Interests}{
The authors have no conflicts of interest to declare that are relevant to the content of this chapter.}

\eject






\bibliographystyle{spphys}
\bibliography{references}

\end{document}